\documentclass[conference]{IEEEtran}
\IEEEoverridecommandlockouts
\usepackage{cite}
\usepackage{verbatim}
\usepackage{amsmath,amssymb,amsfonts}
\usepackage{hyperref}
\usepackage{cleveref}
\usepackage{tikz}
\usepackage{algorithmic}
\usepackage{graphicx}
\usepackage{subfigure}
\def\BibTeX{{\rm B\kern-.05em{\sc i\kern-.025em b}\kern-.08em
    T\kern-.1667em\lower.7ex\hbox{E}\kern-.125emX}}

\newcommand\copyrighttext{%
  \footnotesize \textcopyright 2022 IEEE. Personal use of this material is permitted.
  Permission from IEEE must be obtained for all other uses, in any current or future 
  media, including reprinting/republishing this material for advertising or promotional purposes, creating new collective works, for resale or redistribution to servers or lists, or reuse of any copyrighted component of this work in other works.}
\newcommand\copyrightnotice{%
\begin{tikzpicture}[remember picture,overlay]
\node[anchor=south,yshift=10pt] at (current page.south) {\fbox{\parbox{\dimexpr\textwidth-\fboxsep-\fboxrule\relax}{\copyrighttext}}};
\end{tikzpicture}%
}    

\begin{document}

\title{Analysis and Performance Evaluation of Mobility for Multi-Panel User Equipment in 5G Networks \\
}

\author{\IEEEauthorblockN{Subhyal Bin Iqbal$^{*\dagger}$, Ahmad Awada$^{*}$, Umur Karabulut$^{*}$, Ingo Viering$^{*}$, Philipp Schulz$^{\dagger}$ and Gerhard P. Fettweis$^{\dagger}$}
\IEEEauthorblockA{$^{*}$Standardization and Research Lab, Nokia, Munich, Germany} \
{$^{\dagger}$Vodafone Chair for Mobile Communications Systems, Technische Universität Dresden, Germany}\\
}

\maketitle

%
\copyrightnotice

\begin{abstract}
Frequency Range 2 (FR2) has become an integral part of 5G networks to fulfill the ever increasing demand for user data throughput. However, radio signals in FR2 experience high path and diffraction loss in mobile environments. To address this issue, multi-panel user equipment (MPUE) is adopted for FR2 whereby multiple antenna panels are placed on the UE body to leverage gains from antenna directivity. In contrast to  traditional UEs with isotropic radiation patterns, signal measurements of cells in the network may not be available on all panels simultaneously for MPUE, which may result in outdated signal measurements that affect the reliability of  mobility decisions. In this paper, we investigate the mobility performance of two different MPUE schemes following different paradigms for signal measurement and compare their performance with traditional UEs. This performance evaluation is based in multi-beam 5G networks operating in FR2 where there are multiple simultaneously active beams per cell to realize the high throughput requirements. Furthermore, an in-depth analysis of the mobility performance is carried out to determine the best mobility parameter combinations for the different MPUE schemes. Results have shown that both MPUE schemes offer considerable mobility performance gains as compared to traditional UEs. Moreover, it is seen that the MPUE schemes require different mobility parameter settings for the best mobility performance.
\end{abstract}

\begin{IEEEkeywords}
FR2, 5G, multi-panel UE, signal measurements, performance evaluation, multi-beam networks, mobility parameter settings. 
\end{IEEEkeywords}

\section{Introduction}  \label{Sec1}

Although frequency range 2 (FR2) \cite{b02} addresses the problem of contiguous bandwidth that is needed for 5G networks to fulfill the steep increase in user demand, it brings additional challenges to the link budget such as higher free-space path loss and penetration loss in mobile environments\cite{b03}. One possible solution to address this problem is through the use of multi-panel user equipment (MPUE), i.e., a UE equipped with more than one antenna panel \cite{b04}. There are two benefits that can be availed from the use of MPUE. Firstly, MPUE offers a higher directional antenna gain when compared with a UE that has an isotropic radiation pattern. Secondly, interference from the neighboring cells can be suppressed and a good link quality can be ensured if the MPUE can be made to communicate with one or more panels that are pointing towards the boresight of the transmit beam of the serving cell. However, the procedure to perform signal measurements needed for mobility is now more complex since not all panels of the MPUE are necessarily activated simultaneously \cite{b05, b06}.

In \cite{b06}, the authors have discussed in length the performance gains of MPUE over a reference UE model with isotropic radiation pattern in terms of different mobility key performance indicators (KPIs). This paper investigates the mobility performance in greater detail by jointly analyzing an extended mobility KPI set. A joint analysis of these KPIs is needed to better understand the mobility performance in 5G cellular networks due to the trade-offs involved between the different KPIs \cite{b07}. Thereafter, the extended KPI set is evaluated for different mobility parameter settings for two different MPUE schemes following different paradigms for signal measurement \cite{b08}. The aim is to investigate whether different parameter settings are needed for different MPUE schemes and if so, to determine the best parameter combination that maximizes performance in terms of the extended KPI set. In contrast to \cite{b06}, our network model has the capability to take into account system-level simulations for more than one simultaneously scheduled beams, which are needed to fulfill the high throughput requirements in 5G networks.

This paper uses the low complexity channel model proposed in \cite{b09} to effectively capture the spatial and temporal characteristics of the wireless channel in 5G networks. Based on \cite{b09, b10}, we consider a multi-beam network with transmitter (Tx)-side beamforming and intra-cell beam management procedures. Based on \cite{b11}, the average downlink signal-to-interference-plus-noise ratio (SINR) model for multi-beam networks is considered, which can be used as an input for the radio link failure detection model. The simulation setup is then used for performance evaluation for both the reference UE model and the MPUE model.

The rest of the paper is organized as follows. In \Cref{Sec2} we provide insights into the handover (HO) model and beam management procedures as part of our network model. In \Cref{Sec3}, we discuss different aspects of the MPUE system model, including the different schemes for carrying out signal measurements for mobility. In \Cref{Sec4} we explain the simulation scenario in the performance evaluation. In \Cref{Sec5}, the mobility KPIs are presented and the mobility performance of the two different MPUE schemes are analyzed. Finally, in \Cref{Sec6} we conclude the paper and provide an outlook for future enhancements and extensions.

\section{Network Model} \label{Sec2}

In this section, the HO and intra-cell beam management procedures that form part of the network model are reviewed.

\subsection{HO Model} \label{Subsec2.1}

In order to ensure a reliable HO from the serving cell to the target cell, it is essential that the physical (PHY) layer reference signal received powers (RSRPs) undergo filtering to mitigate the effect of channel impairments. The HO model that is considered in this paper is the baseline HO mechanism of 3GPP Release 15 \cite{b02, b12}. Each UE in the network is capable of measuring the raw RSRP measurements $P_{c,b}^\textrm{RSRP}(n)$ (in dBm) at a discrete time instant $n$ from each beam $b \in B$ of cell $c \in C$, using the synchronization signal block (SSB) bursts that are transmitted by the base station (BS). The separation between the time instants is denoted by $\Delta t$ ms. Layer 1 (L1) and Layer 3 (L3) filtering are applied sequentially by the UE to the raw RSRPs to mitigate the effects of fast-fading and measurement error and calculate the L3 cell quality of the serving and neighboring cells. L1 filtering implementation is not specified in 3GPP standardization and is UE specific. Herein, we use a moving average filter for L1 filtering, where the L1 filter output is expressed as
\vspace{-1mm}
\begin{equation}
\label{Eq1}
    P_{c,b}^\textrm{L1}(m) = \frac{1}{N_\mathrm{L1}} \sum_{i=0}^{N_\mathrm{L1}-1}P_{c,b}^\textrm{RSRP}(m-\omega i), \ m = n\omega 
\end{equation}
where $\omega \in \mathbb{N}$ is the L1 measurement period (aligned with the SSB periodicity) that is normalized by the time step duration $\Delta t$, and $N_\mathrm{L1}$ is the number of samples that are averaged in each L1 measurement period. L1 measurements are then used for cell quality derivation, where we first consider the set $B_{\mathrm{str},c}$ of strongest beams with signal measurements that are above a certain threshold $P_{\mathrm{thr}}$. $B_{\mathrm{str},c}$ is, thus, defined as
\vspace{-1mm}
\begin{equation}
\label{Eq2}
    B_{\mathrm{str},c}(m) = \big\{b \  | \ P_{c,b}^\textrm{L1}(m) > P_{\mathrm{thr}} \big\}.
\end{equation}

Next, $N_\mathrm{str}$ beams representing the subset $B_{\mathrm{str},c}^{\prime}$ of $B_{\mathrm{str},c}$ with the strongest  $P_{c,b}^\textrm{L1}(m)$ are taken and averaged to derive the L1 cell quality of cell $c$ as 
\vspace{-1mm}
\begin{equation}
\label{Eq4}
    P_{c}^\textrm{L1}(m) = \frac{1}{|B_{\mathrm{str},c}^{\prime}|} \sum_{b \in B_{\mathrm{str},c}^{\prime}}  P_{c,b}^\textrm{L1}(m).
\end{equation}

The cardinality of the set is denoted by $|.|$ and the set $B_{\mathrm{str},c}(m)$ is taken as $B_{\mathrm{str},c}^{\prime}$ if $|B_{\mathrm{str},c}(m)| < N_\mathrm{str}$. In case the set $B_{\mathrm{str},c}(m)$ is empty, the L1 cell quality $P_{c}^\textrm{L1}(m)$ is taken as the highest $P_{c,b}^\textrm{L1}(m)$.

Next, the L1 cell quality is further smoothed by L3 filtering to yield the L3 cell quality. We use an infinite impulse response (IIR) filter for L3 filtering, where the L3 filter output is expressed as
\vspace{-1mm}
\begin{equation}
\label{Eq4}
    P_{c}^\textrm{L3}(m) = \alpha_c  P_{c}^\textrm{L1}(m) + (1-\alpha_c)P_{c}^\textrm{L3}(m-\omega),
\end{equation}
where $\alpha_c = (\frac{1}{2})^{\frac{k}{4}}$ is now the forgetting factor controlling the impact of older L3 cell quality measurements $P_{c}^\textrm{L3}(m-\omega)$ and $k$ is the filter coefficient of the IIR filter \cite{b02}.

In the same manner, the L1 RSRP beam measurement $P_{c,b}^\textrm{L1}$ of each beam $b$ of cell $c$ also undergoes L3 filtering, where the output is now the L3 beam measurement $P_{c,b}^\textrm{L3}$ 
\vspace{-1mm}
\begin{equation}
\label{Eq5}
    P_{c,b}^\textrm{L3}(m) = \alpha_b  P_{c,b}^\textrm{L1}(m) + (1-\alpha_b)P_{c}^\textrm{L3}(m-\omega),
\end{equation}
where $\alpha_b$ can be configured independently of $\alpha_c$.

L3 cell quality $P_{c}^\textrm{L3}(m)$ is a measure of the average downlink signal strength for a link between a cell $c$ and a UE and it is used by the network to trigger the HO from the serving cell $c_0$ to one of its neighboring cells $c_\mathrm{N}$, termed as the target cell $c^{\prime}$ \cite{b09}. For intra-frequency HO decisions, typically the A3 trigger condition is configured for measurement reporting\cite{b02}. The UE is triggered to report the L3 cell quality measurement of the target cell $P_{c^{\prime}}^\textrm{L3}(m)$ and L3 beam measurements $P_{c^{\prime},b}^\textrm{L3}(m)$ to its serving cell $c_0$ when the A3 trigger condition, i.e.,
\begin{equation}
\label{Eq6}
     P_{c_0}^\textrm{L3}(m) + o^\mathrm{A3}_c < P_{c^{\prime}}^\textrm{L3}(m) \ \text{for} \  m_0 - T_\mathrm{TTT,A3} < m < m_0,
\end{equation}
expires at the time instant $m=m_0$ for $c^{\prime} \neq c_0$, where $o^\mathrm{A3}_c$ is termed as the HO offset and the observation period in (\ref{Eq6}) is termed as the time-to-trigger $T_\mathrm{TTT,A3}$ (in ms).

After receiving the L3 cell quality measurements, the serving cell $c_0$ sends a HO request to the target cell $c^{\prime}$, which is typically the strongest cell, along with the L3 beam measurements $P_{c^{\prime},b}^\textrm{L3}(m)$ of the target cell $c^{\prime}$. Thereafter, the target cell prepares contention free random-access (CFRA) resources for beams $b \in B_{\mathrm{prep},c^{\prime}}$, e.g., with the highest signal power based on the reported L3 beam measurements. The target cell acknowledges and provides a HO command to the UE. The serving cell forwards the HO command to the UE and once the UE receives this message, it detaches from its serving cell $c_0$ and initiates random-access towards the target cell $c^{\prime}$ using the CFRA resources.

\subsection{Intra-cell Beam Management Procedures} \label{Subsec2.2}
Intra-cell beam management has been defined in 3GPP Release 15 as a set of procedures for the determination and update of serving beam(s) for each UE within a serving cell $c_0$ in the network \cite{b13}. Beam switching is one of these key procedures whereby each UE, after performing L1 filtering of the raw RSRPs of the serving cell $c_0$, reports the $N_\textrm{rep}$ highest L1 beam measurements $P_{c_0,b}$ to the network in a periodic manner. The serving cell $c_0$ then additionally uses L2 IIR filtering on these $N_\textrm{rep}$ reported L1 beam measurements to decide the serving beam $b_0$ for the UE \cite{b10}. If there is a beam among the reported $N_\textrm{rep}$ beams with a L2 RSRP measurement $P_{c_0,b}^\textrm{L2}$ that is beam switching offset $o_b$ (in dB) higher than the serving beam, the BS initiates a beam switch command and informs the UE to switch to the new beam.

The other key procedure of beam management is beam failure detection (BFD). The UE keeps track of the radio link quality (RLQ) of the serving cell $c_0$ by further averaging the instantaneous downlink SINR measurements of the serving cell $\gamma_{c_0,b_0}$  at every discrete simulation time step $n$ to derive the RLQ SINR metric $\bar{\gamma}_\mathrm{RLQ}$ \cite{b14}. The medium access control (MAC) layer of the UE has a beam failure indication (BFI) counter $C_\mathrm{BFI}$, and a BFD timer with $T_\mathrm{BFD}$ time duration. If the RLQ SINR $\bar{\gamma}_\mathrm{RLQ}$ drops below the SINR threshold $\gamma_\mathrm{out}$ an out-of-sync indication is reported to the upper MAC layer, the counter $C_\mathrm{BFI}$ is increased by one, and the timer $T_\mathrm{BFD}$ is started. If the timer expires without a new arrival of an out-of-sync indication, the counter is reset to 0. If another out-of-sync indication is received while the timer is still running, the counter is increased by one and the timer is restarted. A beam failure is detected if the counter reaches a maximum value $C_\mathrm{BFI}^\mathrm{max}$. This prompts the UE to initiate a beam failure recovery (BFR) procedure. The UE first compares the L1 RSRP beam measurements $P_{c_0,b}^\textrm{L1}(m)$ of the serving cell $c_0$ and chooses the beam with the highest L1 RSRP measurement as the target beam $b^{\prime}$. The UE then
attempts random-access on the selected beam and waits for the BS to send a random-access response (RAR) indicating that the access was successful. If the first attempt is unsuccessful, the UE attempts another random-access on the target beam $b^{\prime}$. In total, $N_\mathrm{RACH}$ such attempts are made at time intervals of $T_\mathrm{RACH}$.
If all such attempts are unsuccessful, a radio link failure (RLF) is declared.

\section{MPUE System Model}  \label{Sec3}
\label{Subsection 2.1}

A proper system model is needed for MPUE in order to quantify the mobility performance of MPUE. We divide the MPUE system model into three distinct parts, each of which is further discussed below.

\subsection{MPUE Antenna Design}

In our implementation, we consider the \textit{edge} MPUE design with three directional antenna panels \cite{b05, b13}. Each directional panel is assumed to have a single antenna element. The antenna element power radiation pattern is based on \cite{b13}, where the maximum directional antenna element gain experienced at the boresight is 5 dBi, the elevation $\theta$ and azimuth $\phi$ half power beamwidth (HPBW) are both 90$^{\circ}$, and the backward attenuation is 25 dB. For benchmarking, we consider a reference UE model with a single antenna element that has an isotropic radiation pattern with a gain of 0 dBi \cite{b09}.

\subsection{MPUE Geometric Model}
\label{Subsection 2.2}

\begin{figure} [!t]
\centering
\includegraphics[width = 0.98\columnwidth, height = 5cm]{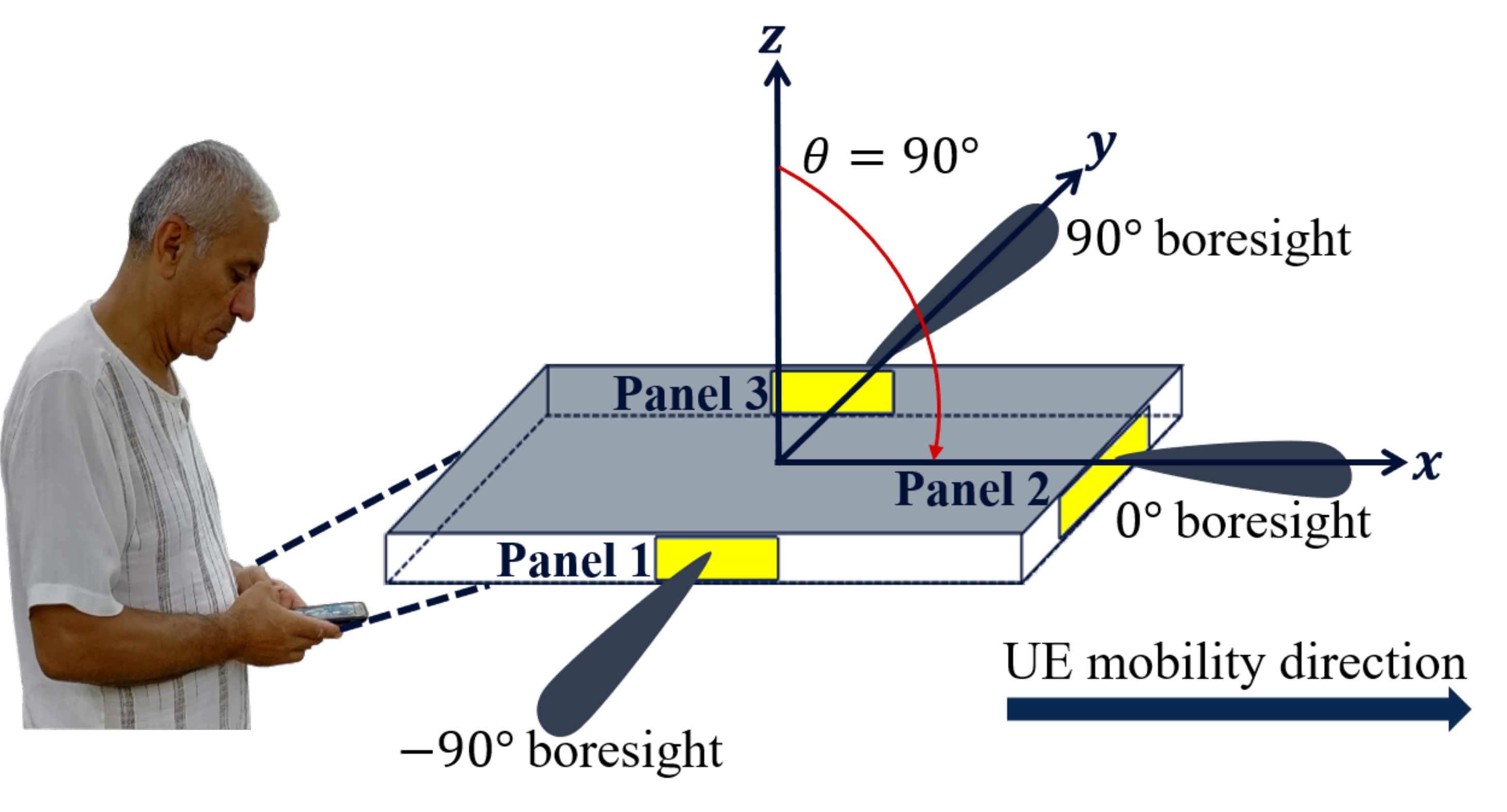}
\caption{An illustration of the MPUE geometric model, where the UE screen is assumed parallel to the ground and panel 2 is always oriented along the UE mobility direction.}
\label{Fig1}
\end{figure}

One of the main challenges in the MPUE system model is determining the directional antenna gain of the different panels for MPUEs as they change position with respect to the boresights of the serving and neighboring BSs. The geometric model we consider accurately models and calculates the directional antenna gain for the different panels, which we incorporate into the link budget design in order to leverage MPUE gains from panel directivity. There are three key assumptions that are considered in our MPUE geometric model, as shown in Fig. \ref{Fig1}:
\begin{itemize}
    \item 
    \textit{UE orientation}: The UE screen, held by a user, is assumed to be parallel to the ground.
    \item 
    \textit{Coordinate system}: A Cartesian coordinate system is assumed, where the elevation angle $\theta=$ 0$^{\circ}$ points to the zenith and $\theta=$ 90$^{\circ}$ points to the horizon.
    \item 
    \textit{Panel geometry}: The boresight of panel 1 (P1), panel 2 (P2) and panel 3 (P3) are assumed at azimuth $\phi$ angles of -90$^{\circ}$, 0$^{\circ}$ and 90$^{\circ}$ respectively in the Cartesian coordinate system. This implies that P2 is always oriented along the UE mobility direction, which is the same as the user mobility direction.
\end{itemize}

\vspace{-2mm}

\subsection{MPUE Schemes for Signal Measurement}
\label{Subsection 2.3}

In line with \cite{b08}, we consider two MPUE schemes for performing signal measurements and introduce the concept of serving and best panel for determining the panels that shall be used in measurement reporting for beam management and HO procedures, respectively. The two MPUE schemes are:
\begin{itemize}
\item 
\textit{MPUE-Assumption 3 (MPUE-A3)}: A UE can measure the RSRPs from its serving cell $c_0$ and neighboring cells $c_\mathrm{N}$ on all three panels simultaneously.
\item
\textit{MPUE-Assumption 1 (MPUE-A1)}: A UE can measure the RSRPs from its serving cell $c_0$ and neighboring cells $c_\mathrm{N}$ on just one panel that is active at a time.
\end{itemize}

In MPUE-A3, the UE measures the raw beam panel RSRPs $P_{c,b,p}$ using SSBs at discrete time instant $n$ for each beam $b \in B$ of a cell $c \in C$ on each of its panels $p \in P$. However, communication with the serving cell $c_0$ through the serving beam $b_0$ still occurs through one panel, which we denote as the serving panel $p^{\mathrm{S}}$. As discussed in \Cref{Subsec2.1}, the raw beam panel RSRPs undergo L1 filtering. If no panel switching offset $o_p$ is defined, the serving panel $p^{\mathrm{S}}$ is the panel that has the strongest L1 beam panel RSRP $P_{c_0,b_0,p}^\mathrm{L1}(m)$ and is defined as 
\vspace{-1mm}
\begin{equation}
\label{Eq8} 
p^{\mathrm{S}} = \arg \max_{p} P_{c_0,b_0,p}^\mathrm{L1}(m).
\end{equation}

The UE switches to another panel $p$ only if it has a stronger L1 beam panel RSRP for the serving beam $b_0$. This decision is fully UE-centric and is made independent of the network. The serving panel $p^{\mathrm{S}}$ has two key purposes. Firstly, it is used for beam reporting for intra-cell beam management as discussed in \Cref{Subsec2.2}. Secondly, the raw beam panel RSRPs measured on $p^{\mathrm{S}}$ are used for calculating the instantaneous downlink SINR $\gamma_{c,b}$ of a link between the UE and beam $b$ of cell $c$ \cite{b11}. As will be discussed later in \Cref{Sec4}, the SINR $\gamma_{c,b}$ is used for HO and RLF determination.

The best panel $p_c^{\mathrm{E}}$ is selected as the panel with the strongest L1 beam panel measurement $P_{c,b,p}(m)$ for any beam $b$ of cell $c$ in the network and is defined as
\vspace{-1mm}
\begin{equation}
\label{Eq9} 
p_c^{\mathrm{E}} = \arg \max_{b,p} P_{c,b,p}^{\mathrm{L1}}(m).
\end{equation}

\begin{figure}[!t]
\centering
\includegraphics[width = 0.96\columnwidth, height=6.35cm]{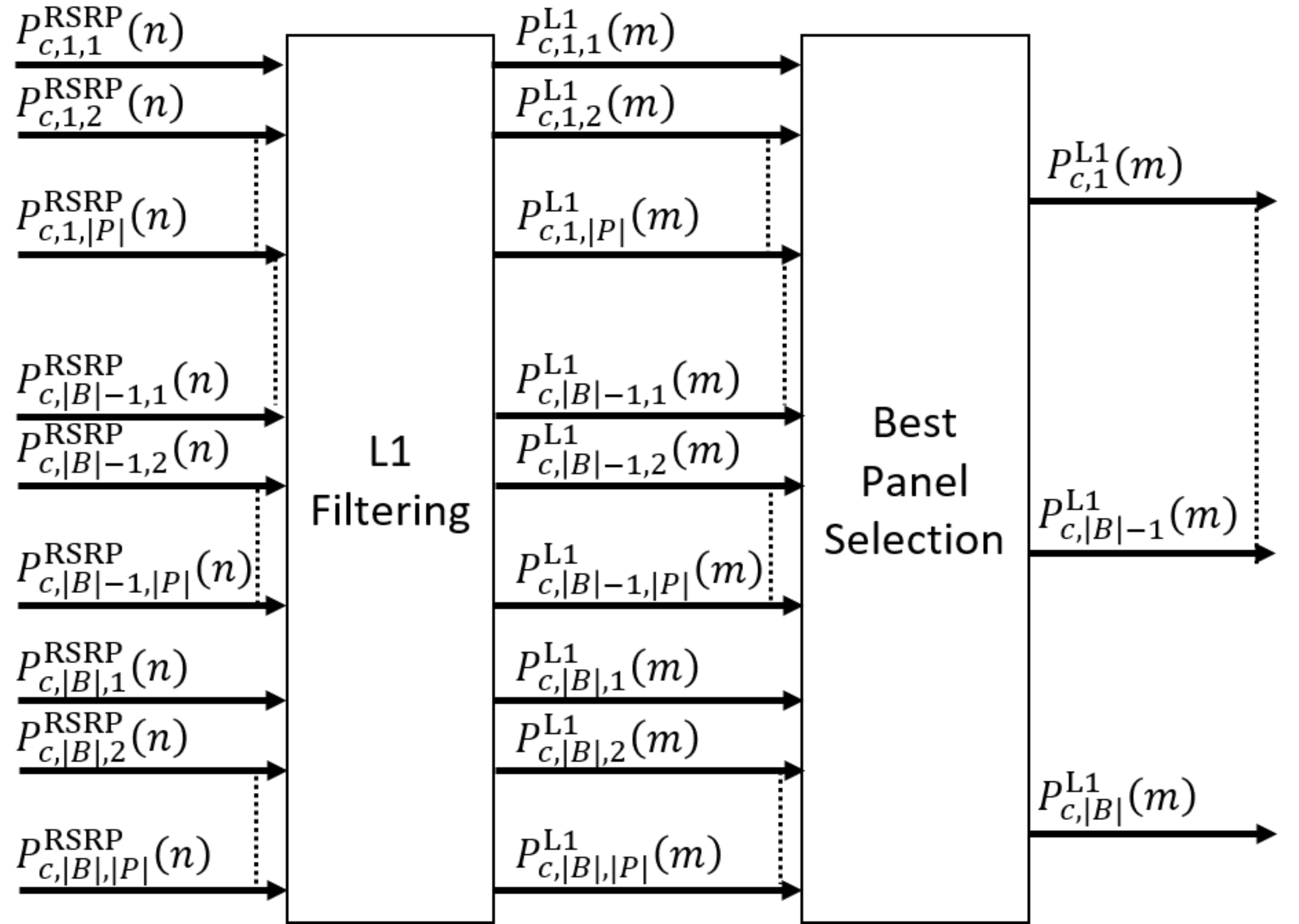}
\caption{L1 filtering and best panel selection to determine the L1 beam RSRPs that are then further used to determine the L3 cell quality measurement and L3 beam quality measurements.}
\label{Fig2}
\end{figure}

The L1 beam panel RSRPs $P_{c,b,p}^{\mathrm{L1}}(m)$ of the best panel $p_c^{\mathrm{E}}$ are denoted as L1 beam RSRPs $P_{c,b}^{\mathrm{L1}}(m)$ and are used for deriving the L3 cell quality measurement $P_{c}^{\mathrm{L3}}(m)$ and L3 beam measurements $P_{c,b}^{\mathrm{L3}}(m)$, as explained in \Cref{Subsec2.1}. This is illustrated in Fig. \ref{Fig2}. The L3 cell quality measurements are then used for HO decisions.

\begin{figure}[!b]
\centering
\includegraphics[width = 0.91\columnwidth]{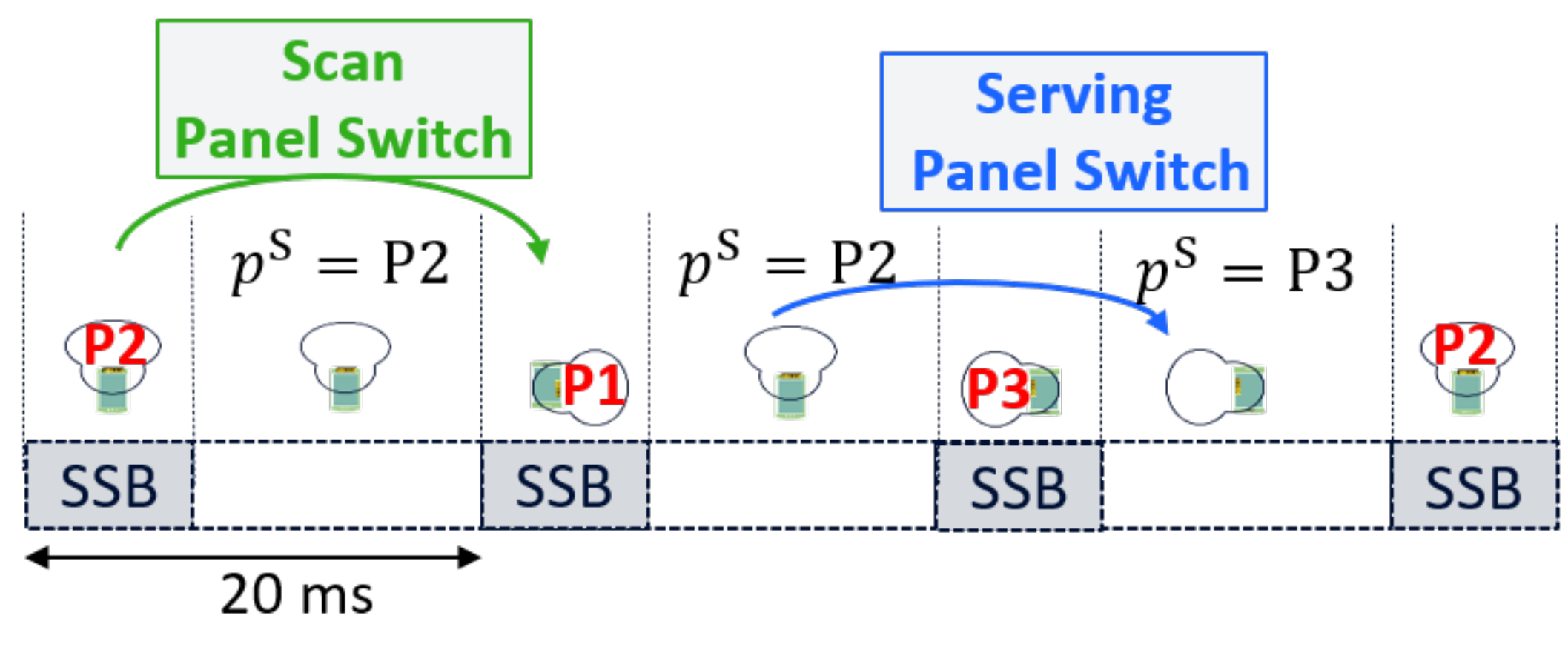}
\caption{Panel scanning and serving panel switch mechanism in MPUE-A1.}
\label{Fig3}
\end{figure}

In contrast, in MPUE-A1 the UE is capable of measuring only on one panel that is active at a given time. The serving panel is still determined as per \eqref{Eq8} and the UE switches to any other panel if the serving beam $b_0$ has a higher L1 RSRP on another panel other than the serving panel $p^\mathrm{S}$. It is left for the UE implementation on how the measurements are to be taken on the active panel. In our implementation, we choose to scan the panels in a round-robin pattern every 20 ms, i.e., the SSB periodicity in a time-multiplexing manner. As seen in Fig. \ref{Fig3}, the UE scans the raw beam panel RSRPs on P2 when the first SSB is transmitted. As per \eqref{Eq9}, it determines P2 as the serving panel $p^\mathrm{s}$. When the second SSB block is transmitted, the UE scans P1 but P2 is still the serving panel because it has a higher L1 RSRP for the serving beam $b_0$ of the serving cell $c_0$. In the third SSB, the UE scans P3 and decides that P3 is the serving panel $p^\mathrm{S}$ and therefore switches its serving panel to P3. Finally, in the fourth SSB the round-robin pattern is complete and the UE is now back to scan P2.

\section{Simulation Scenario and Parameters}  \label{Sec4}

In this section, the simulation scenario is described along with the simulation parameters which are listed in \Cref{Table1}.

\begin{table}[!b]
\renewcommand{\arraystretch}{1.3}
\caption{SIMULATION PARAMETERS}
\centering
\begin{tabular}{l l}
\hline
\bfseries Parameter & \bfseries Value\\
\hline\hline
Carrier frequency & 28 GHz\\
System bandwidth & 100 MHz\\
Cell deployment topology & 7-site hexagon\\
Total number of cell $N_\mathrm{cells}$ & 21 \\
Downlink Tx power & 40 dBm\\
Tx (BS) antenna height & 10 m\\
Tx antenna element pattern & Table 7.3-1 in \cite{b15} \\
Tx panel size & 16 $\times$ 8, $\forall b \in \{1,\ldots,8\}$\\
 & 8 $\times$ 4, $\forall b \in \{9,\ldots,12\}$ \\
Tx vertical antenna element spacing & 0.7$\lambda$ \\
Tx horizontal antenna element spacing & 0.5$\lambda$ \\
Beam elevation angle $\theta_b$ & 90, $\forall b \in \{1,\ldots,8\}$ \\
 & 97, $\forall b \in \{9,\ldots,12\}$\\
Beam azimuth angle $\phi_b$ & $-$52.5$+$15$(b-1), \forall b \in \{1,\ldots,8\}$\\
& $-$45$+$30$(b-9), \forall b \in \{9,\ldots,12\}$\\
Beamforming gain model & Fitting model of \cite{b09}\\
Rx (UE) antenna height & 1.5 m\\
Rx antenna element pattern & Reference UE model: isotropic,\\ 
 & MPUE model: based on \cite{b13} \\
Rx panel size  & Single antenna element\\
Rx antenna element antenna gain  & Reference model: 0 dBi\\ 
& MPUE model: 5 dBi\\
Total number of UEs $N_\mathrm{UE}$ & 420\\
UE speed  & 30 km/h\\
UE panel switching offset $o_p$ & 0 dB\\
Penetration loss & 0 dB\\
Fast-fading channel model & Abstract model of \cite{b09}\\
Time step $\Delta t$  & 10 ms\\
SSB periodicity (beam sweeping period) & 20 ms\\
Normalized L1 measurement period $\omega$ & 2\\
L1 filter length $N_\mathrm{L1}$ & 2\\
SINR threshold $\gamma_\mathrm{out}$  & $-$8 dB \\
\hline
\end{tabular}
\label{Table1}
\end{table}

We consider a 5G network model with an urban-micro (UMi) cellular deployment consisting of a standard hexagonal grid with 7 BS sites, each divided into three sectors or cells. The inter-cell distance is 200 meters and the carrier frequency is 28 GHz. In total, 420 UEs are dropped randomly with 2D uniform distribution over the network at the beginning of simulation, each moving at a constant velocity of 30 km/h in random waypoint motion along straight lines. For the MPUE model, the UE panel switching offset $o_p$ is assumed to be 0 dB. Moreover, no delay is assumed when the UE switches between serving panels.

\begin{figure}[!b]
\textit{\centering
\includegraphics[width = 0.96\columnwidth]{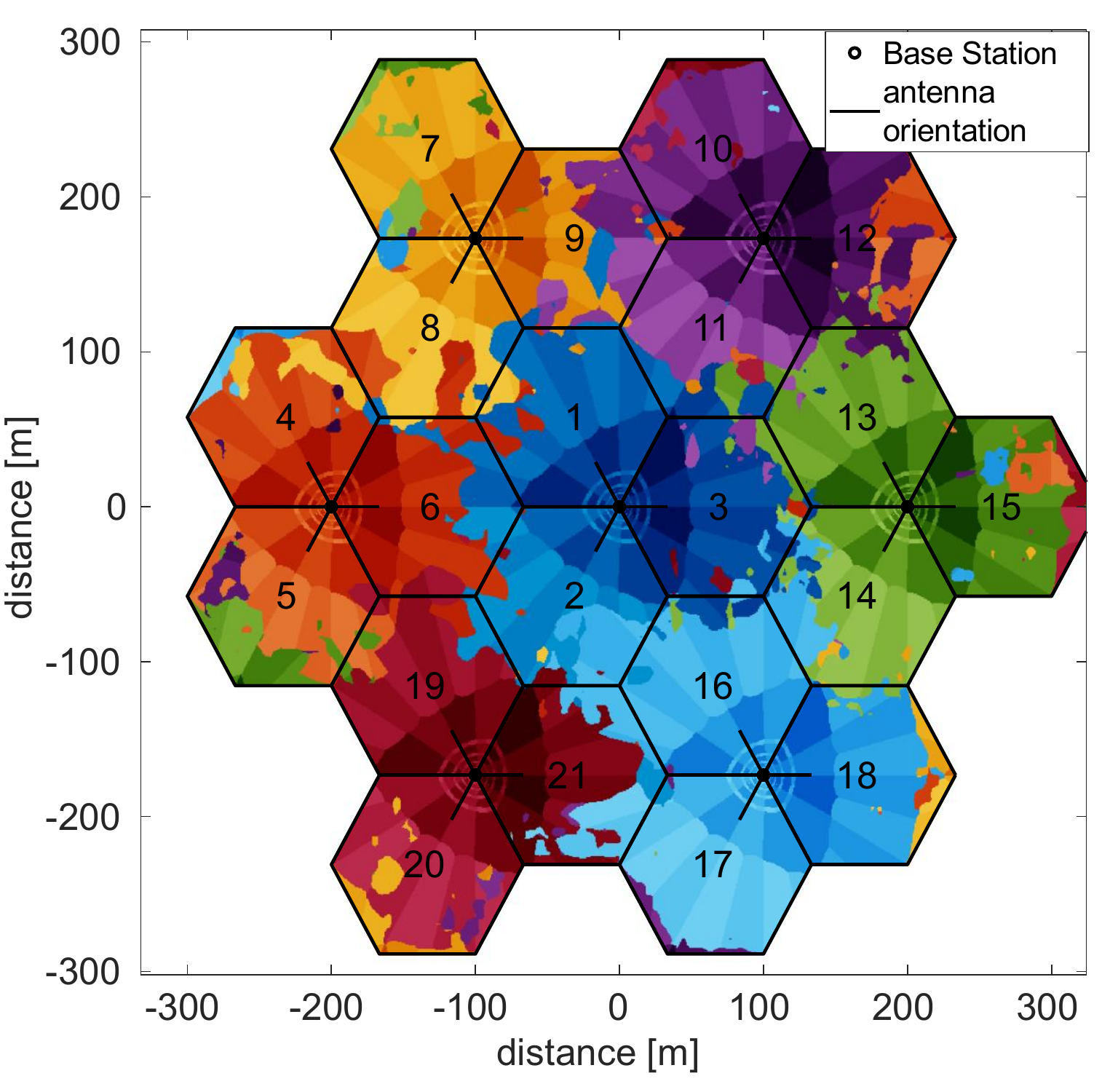}
\vspace{-\baselineskip}
\caption{Simulation scenario consisting of seven hexagonal sites, where each site is serving 3 cells with 120$^{\circ}$ coverage. Tx side beamforming is considered, consisting of 12 beams in each cell.} 
\label{Fig4}}
\end{figure}

As per \cite{b15}, shadow fading due to large obstacles is taken into account and all radio links between the BSs and the UEs are assumed to have soft line-of-sight (LOS). The fast-fading channel model used is the low complexity channel model for multi-beam systems proposed in \cite{b09}, which integrates the spatial and temporal characteristics of 3GPP's geometry-based stochastic channel model (GSCM) \cite{b15} into Jake’s channel model. The beamforming gain model is based on \cite{b09}, where a 12-beam grid of beams (GoBs) configuration is considered. This can be observed in Fig. \ref{Fig4}. $K_b=$ 4 beams are simultaneously scheduled  for all cells in the network. Beams $b \in \{1,\ldots,8\}$ have smaller beamwidth and higher beamforming gain and therefore cover regions further apart from the BS (light color). Beams  $b \in \{9,\ldots,12\}$ have larger beamwidth and relatively smaller beamforming gain and cover regions closer to the BS (dark color). The effect of shadow fading is also visible as coverage islands in Fig. \ref{Fig4}.


The instantaneous downlink SINR $\gamma_{c,b}(m)$ of a link between the UE and beam $b$ of cell $c$ is evaluated by the Monte-Carlo approximation given in \cite{b11} for the strict fair resource scheduler. This SINR is of key importance in the HOF and RLF models, each of which are elaborated below.

\textit{HOF Model:} The HOF model is used to model failure of a UE to hand over from its serving cell $c_0$ to its target cell $c^{\prime}$. As discussed in \Cref{Subsec2.1}, a UE initiates a HO by using the CFRA resources to access the selected beam $b^{\prime}$ of target cell $c^{\prime}$. For successful random-access, it is a prerequisite that the SINR $\gamma_{c^{\prime},b^{\prime}}(m)$ of the target cell remains above the threshold $\gamma_\mathrm{out}$ during the RACH procedure. A HOF timer $T_\mathrm{HOF} $ = 200 ms is started when the UE initiates the random-access towards the target cell $c^{\prime}$. A UE only succeeds in accessing the target cell if the SINR  $\gamma_{c^{\prime},b^{\prime}}(m)$ remains above the threshold $\gamma_\mathrm{out}$ while the timer $T_\mathrm{HOF}$ is running, and as such a successful HO is declared. A HOF is declared if the timer $T_\mathrm{HOF}$ expires and the UE fails to access the target cell, i.e., $\gamma_{c^{\prime},b^{\prime}}(m)<\gamma_\mathrm{out}$. The UE then performs connection re-establishment to a new cell (possibly the previous serving cell) and this procedure contributes to additional signaling overhead and latency \cite{b02}. 

\textit{RLF Model:} The RLF model is used to model failure of a UE in its serving cell $c_0$. The UE keeps track of the radio link monitoring (RLM) SINR metric $\bar{\gamma}_\mathrm{RLM}$, which is an average of the instantaneous downlink SINR measurements of the serving cell $\gamma_{c_0,b_0}$. A RLF timer $T_\mathrm{RLF}$ = 1000 ms is started when the RLM SINR $\bar{\gamma}_\mathrm{RLM}$ of the serving cell drops below $\gamma_\mathrm{out}$, and if the timer $T_\mathrm{RLF}$ expires a RLF is declared. The UE then initiates connection re-establishment. While the timer $T_\mathrm{RLF}$ runs, the UE may recover before declaring a RLF if the SINR $\bar{\gamma}_\mathrm{RLM}$ exceeds a second SINR threshold defined as $\gamma_\mathrm{in}$ = $-$6 dB, where $\gamma_\mathrm{in} > \gamma_\mathrm{out}$\cite{b02}. As discussed in \Cref{Subsec2.2}, if the BFR process fails the UE also declares a RLF and this is also taken into account in the RLF model.

\section{Performance Evaluation} \label{Sec5}

In this section, the mobility performance of the two different MPUE schemes are evaluated against that of the reference UE model. Moreover, we analyze whether the two different MPUE schemes require distinct mobility parameter settings. The KPIs used for evaluation are explained below.

\subsection{KPIs}
\label{Sec5.1}

\textit{Successful HO:} Indicates the total number of successful HOs from the serving cell $c_0$ to the target cell $c^{\prime}$ in the network.

\textit{Fast HO}: Indicates the sum of ping-pongs and short-stays in the network. A ping-pong is a successful HO followed by a HO back to the original cell within a very short time $T_{FH}$ \cite{b07}, e.g,. 1 second. It is assumed that both HOs could have been potentially avoided. A short-stay is a HO from one cell to another and then to a third one within $T_{FH}$. Here it is assumed that a direct HO from the first cell to the third one would have served the purpose. Although fast HOs are part of successful HOs, they are accounted for as a detrimental mobility KPI which adds unnecessary signalling overhead to the network.    

\textit{Failure:} Indicates the the sum of the HOFs and RLFs.

Successful HOs, fast HOs and failures are expressed as percentages of the total number of HO attempts in the network, which is taken as the sum of successful HOs and failures.

\textit{Outage:} Outage is defined as a time period when a UE is not able to receive data from the network due a number of reasons. When the instantaneous SINR of the serving cell $\gamma_{c_0, b_0}$ falls below $\gamma_\mathrm{out}$ it is assumed that the UE is not able to communicate with the network and, thus, in outage. Besides, if the HOF timer $T_{\mathrm{HOF}}$ expires due to a HOF or the RLF timer $T_{\mathrm{RLF}}$ expires due to a RLF, the UE initiates connection re-establishment and this is also accounted for as outage. A successful HO, although a necessary mobility procedure, contributes also to outage since the UE can not receive any data during the time duration the UE is performing random access to the target cell $c^{\prime}$. This outage is modeled as relatively smaller than the outage due to connection re-establishment \cite{b07}. Outage is denoted in terms of a percentage as 
\vspace{-1mm}
\begin{equation}
\label{Eq10} 
\textrm{Outage} \ (\%) = \frac{\sum_{\forall u}{\textrm{Outage duration of UE}} \ u} {N_\mathrm{UE} \ \textrm{x} \ \textrm{Simulated  time}} \ \textrm{x} \ 100. 
\end{equation}

 \subsection{Simulation results} 
 \label{Sec5.2}

\begin{figure}[!b]
\textit{\centering
\includegraphics[width = 0.985\columnwidth, height = 7.0cm]{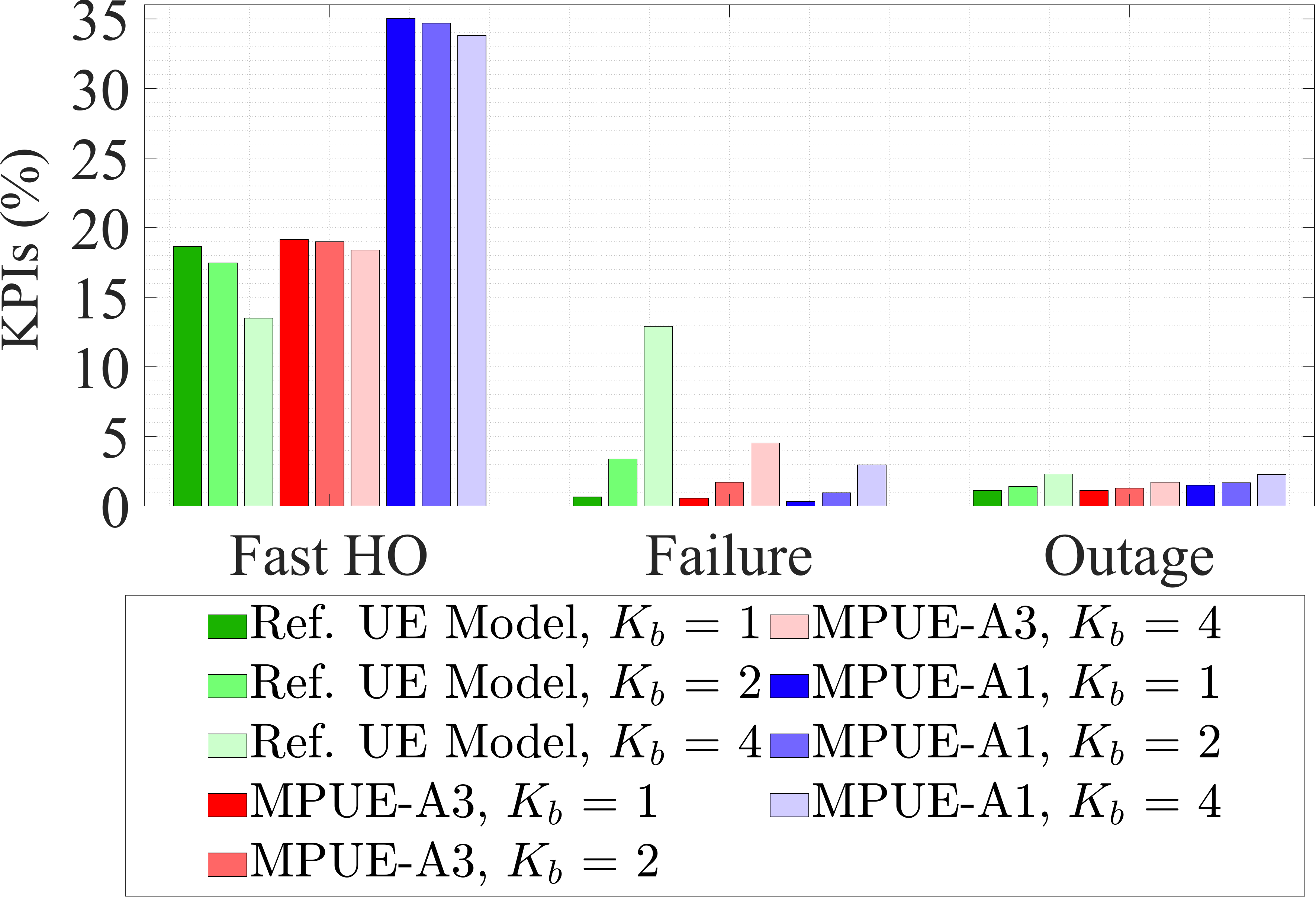}
\caption{Performance analysis of MPUE-A3 and MPUE-A1 for HO offset $o_c^\mathrm{A3}$ = 2 dB and time-to-trigger $T_\mathrm{TTT,A3}$ = 80 ms, where fast HO and failure are expressed as a percentage of total HO attempts in the network and outage is expressed as a percentage as per \eqref{Eq10}.} 
\label{Fig5}}
\end{figure}

 The mobility performance of the reference UE model (shown in green) is compared with MPUE-A3 (shown in red) and MPUE-A1 (shown in blue) for $K_b \in \{1,2,4\}$ in Fig. \ref{Fig5}. The different shades represent the different values of $K_b$ simultaneously scheduled beams per cell.

\begin{figure}
    \centering
    \subfigure[MPUE-A3.]
    {
        \includegraphics[width = 0.96\linewidth]{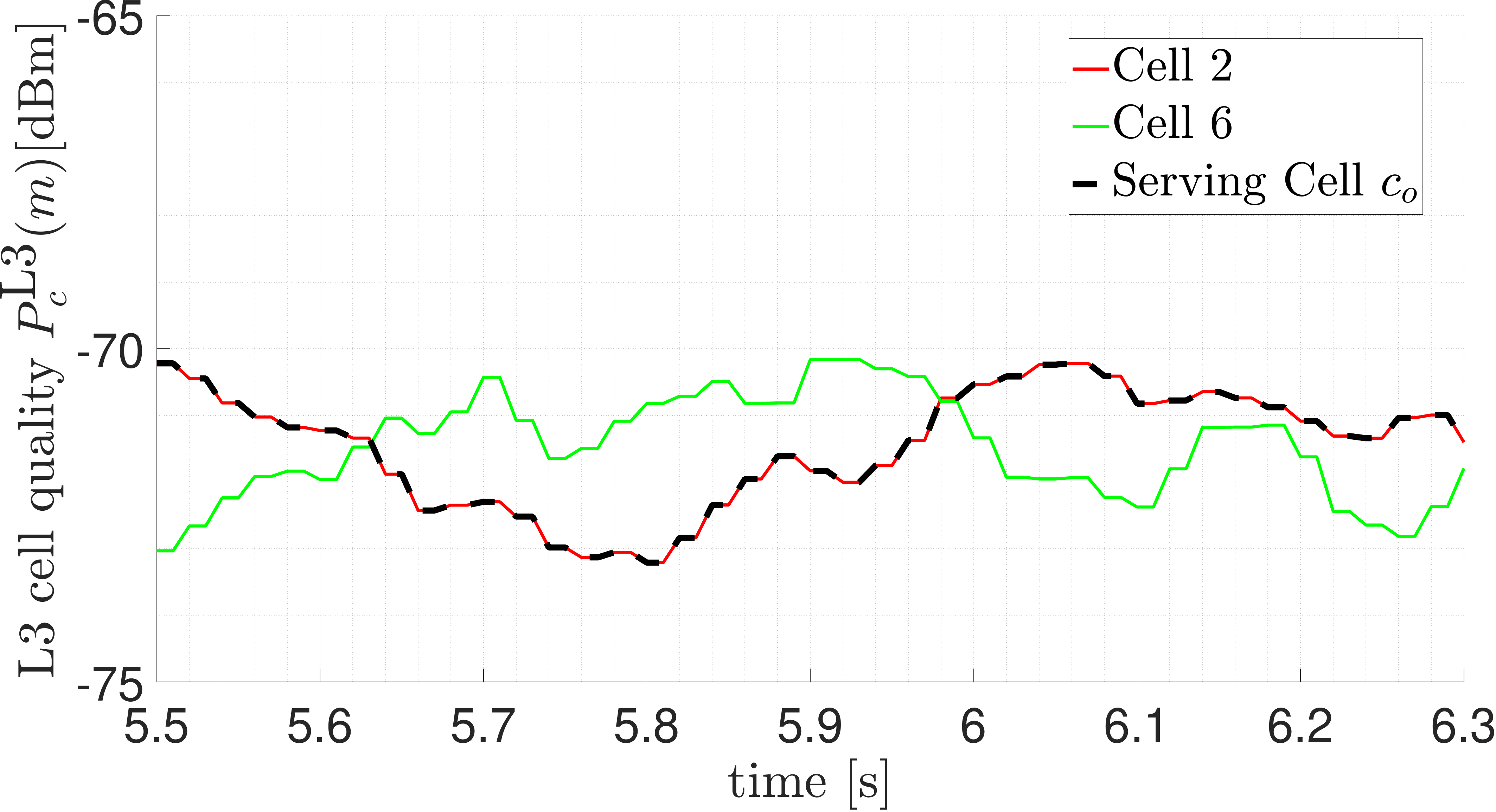}
        \label{fig:Fig7a}
    }
    \\
    \subfigure[MPUE-A1.]
    {
        \includegraphics[width = 0.96\linewidth]{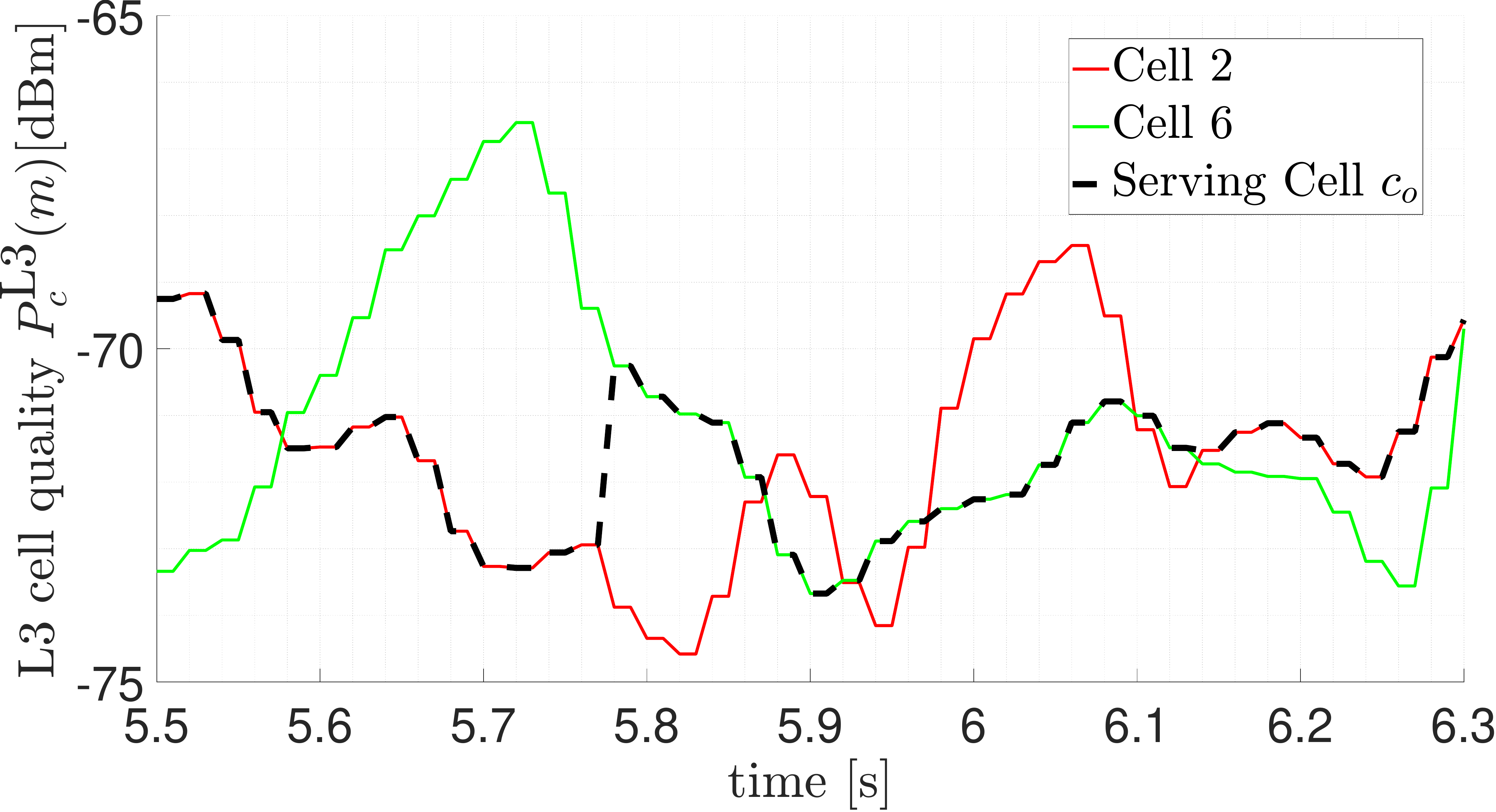}
        \label{fig:Fig7b}
    }
    \caption{L3 cell quality measurements $P_{c}^\textrm{L3}(m)$ for serving cell $c_0$, cell 2 and cell 6 as a function of time [s] for (a) MPUE-A3 and (b) MPUE-A1.}
    \label{fig:Fig7} \vspace{-\baselineskip}
\end{figure}

The first key observation is that both MPUE schemes are more robust to increasing SINR degradation caused by the higher $K_b$ value. As $K_b$ is increased from 1 to 4, the percentage of failures increases by 12.3\% (from 0.7\% to 13.0\%) for the reference UE model. In comparison, the percentage of failures increase by 3.9\% (from 0.6\% to 4.5\%) and 2.7\% (from 0.3\% to 3.0\%) for MPUE-A3 and MPUE-A1, respectively. By comparing fast HOs, we see that the percentage of fast HOs for MPUE-A3 is almost half as that for MPUE-A1. This is due to the fact that for MPUE-A1, following the round-robin approach for signal measurements the RSRP on two out of three panels are outdated, resulting in the UE triggering more unnecessary HOs. This is illustrated in Fig. \ref{fig:Fig7}, showing the L3 cell quality measurements $P_{c}^\textrm{L3}(m)$ for serving cell $c_0$ and two other cells 2 and 6 as a function of time. During the considered evaluation time, the serving cell for MPUE-A3 remains cell 6 while for MPUE-A1 a HO is experienced for cell 2 and 6 at time instant 5.78 and 6.14 seconds, respectively. This can be classified as a ping-pong. The high number of fast HOs in MPUE-A1 has also an inadvertent effect on the radio and network signalling overhead that is needed for HO preparation. 

For the reference UE model, the outage percentage increases by 1.2\% when $K_b$ is increased from 1 to 4 due to the relatively large increase in failures. For MPUE-A3 and MPUE-A1 the outage percentage increases by 0.6\% and 0.8\%, respectively, where the slightly larger outage for MPUE-A1 stems from the relatively high number of fast HOs. In general, it can be said that MPUE-A3 performs better than MPUE-A1 for any $K_b$. For example, for $K_b=4$, although failures are comparable for both MPUE-A3 and MPUE-A1, fast HOs are 15.4\% greater for MPUE-A1.

\begin{figure}
    \centering
    \subfigure[MPUE-A3.]
    {
        \includegraphics[width = 0.96\linewidth]{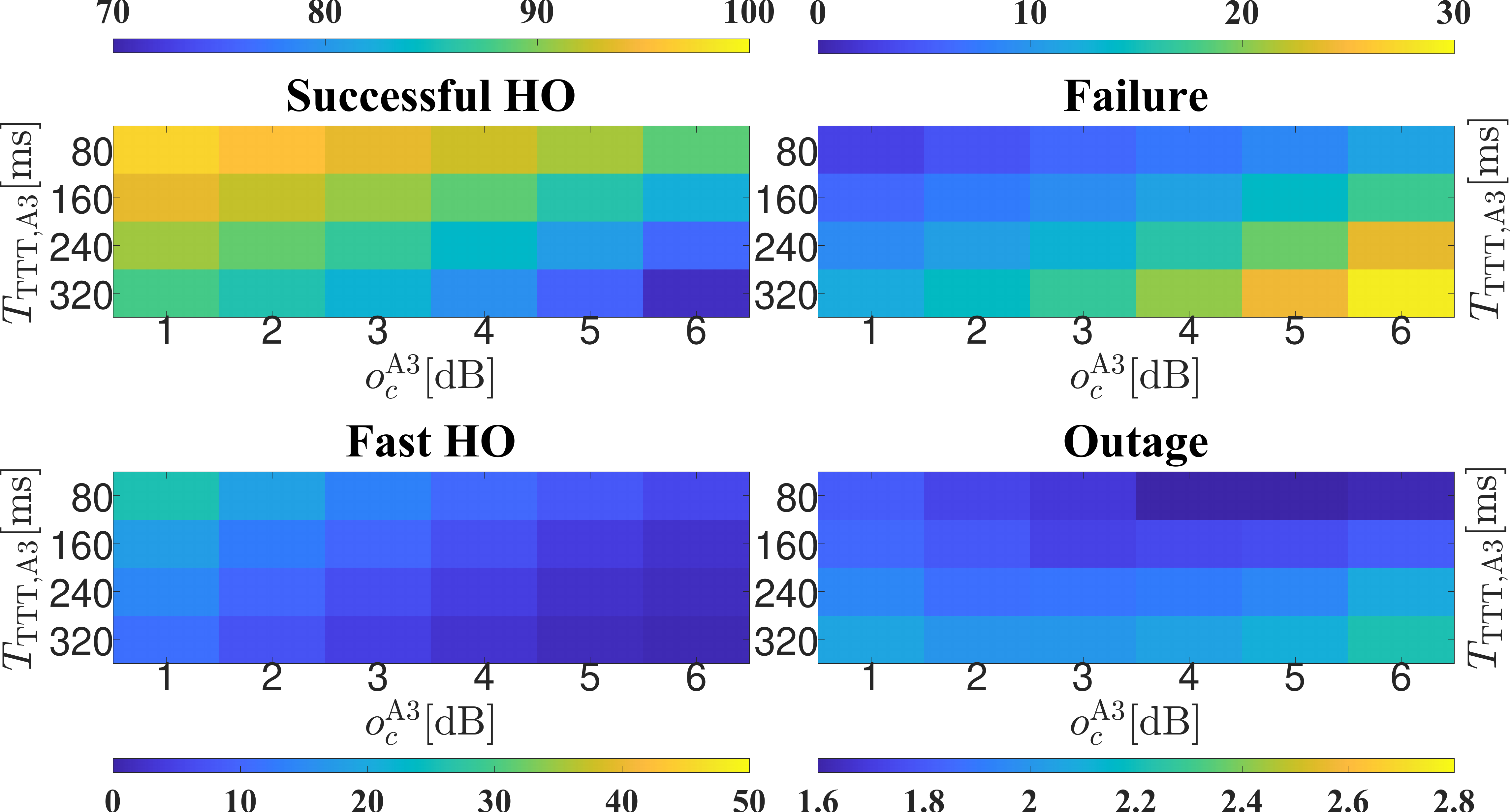}
        \label{fig:Fig6a}
    }
    \\
    \subfigure[MPUE-A1.]
    {
        \includegraphics[width = 0.96\linewidth]{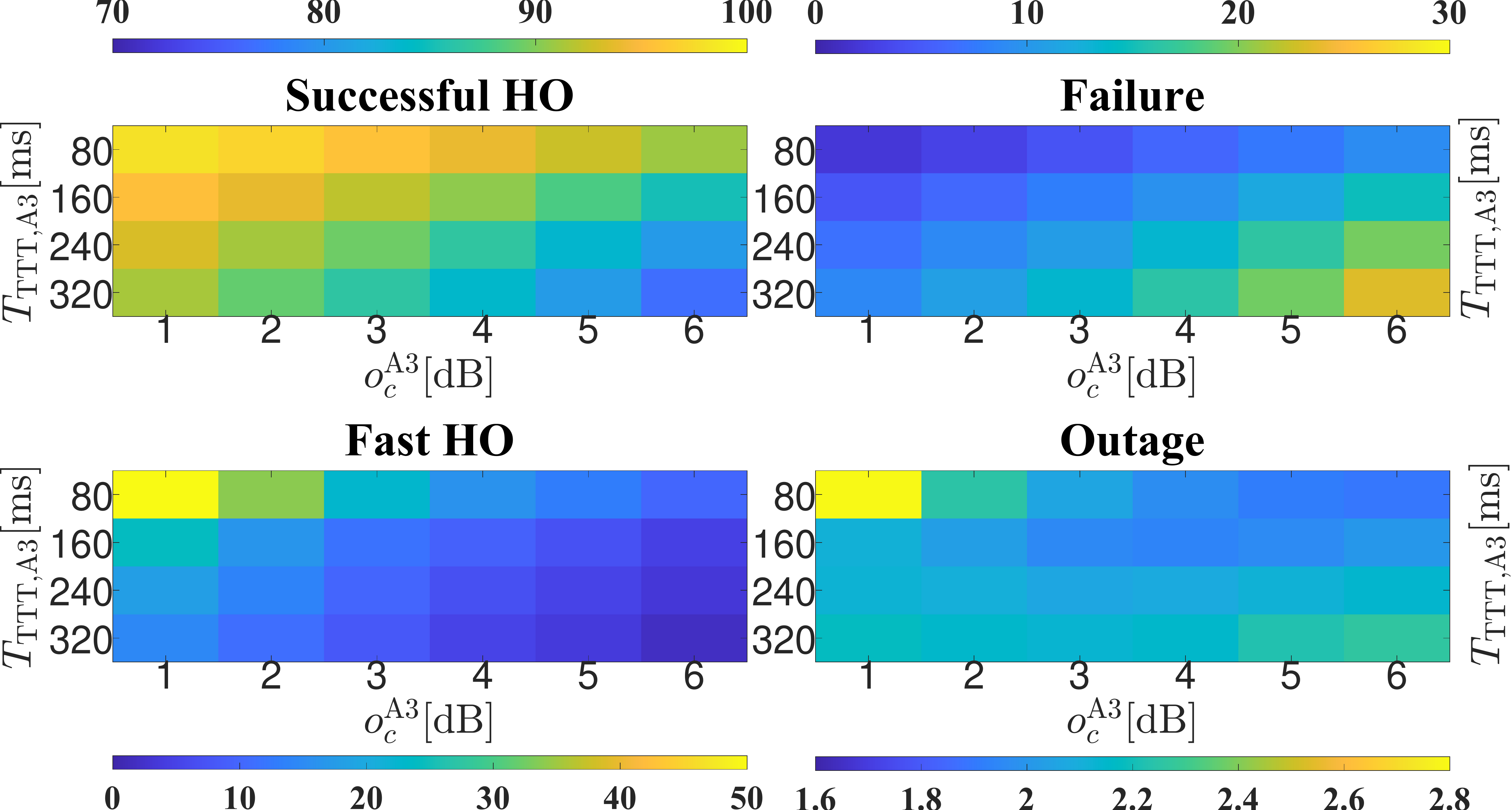}
        \label{fig:Fig6b}
    }
    \caption{Mobility KPIs expressed as percentages for $K_b=4$ in a sweep set consisting of the HO offset $o_c^\mathrm{A3}$ = $\{1,2,3,4,5,6\}$ dB and time-to-trigger $T_\mathrm{TTT,A3}$ = $\{80,160,240,320\}$ ms for (a) MPUE-A3 and (b) MPUE-A1.}
    \label{fig:Fig6} \vspace{-\baselineskip}
\end{figure}

The mobility KPIs for both MPUE-A3 and MPUE-A1 are evaluated for an extended parameter sweep of the HO offset $o_c^\mathrm{A3} \in \{1,2,3,4,5,6\}$ and time-to-trigger $T_\mathrm{TTT,A3} \in \{80,160,240,320\}$. The results are shown in Fig. \ref{fig:Fig6} for successful HOs, failures, fast HOs and outage. As can be seen, there exists a trade off between failures and fast HOs, such that increasing $o_c^\mathrm{A3}$ (while keeping $T_\mathrm{TTT,A3}$ constant) increases the mobility failures while decreases fast HOs. Increasing $T_\mathrm{TTT,A3}$ produces the same effect. From a mobility performance perspective, minimizing failures has a higher priority than minimizing the number of fast HOs since the former leads to higher outage percentage \cite{b07}. The aim is to determine a trade off between $o_c^\mathrm{A3}$ and $T_\mathrm{TTT,A3}$ that gives a maximal number of successful HOs and minimal number of failures, while keeping both fast HOs and outage percentage within a reasonable limit.

For MPUE-A3 in Fig. \ref{fig:Fig6a}, the best combination of the handover offset  and time-to-trigger is determined as $o_c^\mathrm{A3}$ = 2 dB and  $T_\mathrm{TTT,A3}$ = 80 ms, respectively. Failures and fast HOs are 4.5\% and 18.4\% of the total HO attempts, respectively, meaning 95.5\% of the total HO attempts end up as successful HOs. UEs on average spend 1.7\% of their time in outage which is still at the lower end. If we set the parameter combination as $o_c^\mathrm{A3}$ = 1 dB, $T_\mathrm{TTT,A3}$ = 80 ms, it is seen that while failures may decrease by 1.5\% to their minimum value of 3.0\%, fast HOs increase by 7.2\% to more than one-quarter of the total number of the total handover attempts. The trade-off in this case would favor the lower fast HOs values. 

For MPUE-A1 in Fig. \ref{fig:Fig6b} the best parameter combination is determined as $o_c^\mathrm{A3}$ = 3 dB, $T_\mathrm{TTT,A3}$ = 80 ms. The higher offset value $o_c^{\mathrm{A3}}$ of 3 dB is needed to mitigate the exceptionally high percentage of fast HOs compared to MPUE-A3 seen earlier also in Fig. \ref{Fig5}. Failures and fast HOs are  4.4\%  and  22.7\% of  the  total  HO  attempts, respectively, while outage is 2.0\%. Compared to the best parameter combination we have determined for MPUE-A3 in Fig. \ref{fig:Fig6a}, failures are comparable but fast HOs for the best combination are still 4.3\% higher leading to a 0.3\% higher outage.

\section{Conclusion}  \label{Sec6}
\label{Sec6}
In this paper, we have analyzed the performance gain of two MPUE schemes over an isotropic UE (reference model) for different number of $K_b$ simultaneously transmitted beams per cell in 5G multi-beam networks. As higher $K_b$ is needed to realize the high throughput requirements in FR2 \cite{b11}, it is seen that both MPUE schemes are more robust to SINR degradation resulting from higher $K_b$ values when compared with the reference UE model. For $K_b$ = 4 it is observed that failures are 8.5\% and 10.0\% lower for MPUE-A3 and MPUE-A1, respectively, when compared to the reference UE model. Furthermore, it has been shown that different mobility parameter settings are needed for MPUE-A3 and MPUE-A1 to achieve the best mobility performance regarding the mobility parameters values investigated in this study. This is mainly because cell quality measurements in MPUE-A1 are partially outdated compared to those derived in MPUE-A3, leading to many unnecessary HOs such as ping-pongs and short stays. Based on these findings, further studies may utilize UE-group specific mobility robustness optimization techniques to achieve the best mobility performance in the network \cite{b17}.

\begin{thebibliography}{00}

\bibitem{b02} 3GPP, “NR; Radio Resource Control (RRC) Protocol Specification,” Technical Report (TR) 38.331 V16.0.0, 3rd Generation Partnership Project (3GPP), Mar. 2020. 

\bibitem{b03} S. Rangan, T. S. Rappaport, and E. Erkip, “Millimeter-Wave Cellular Wireless Networks: Potentials and Challenges,” \textit{Proceedings of the IEEE}, vol. 102, no. 3, pp. 366–385, 2014.

\bibitem{b04} Qualcomm, “Breaking the Wireless Barriers to Mobilize 5G NR mmWave,”. Qualcomm Webinar, Jan. 2019,  available at :https://www.qualcomm.com/documents/webinar-breaking-wireless-barriers-mobilize-5g-nr-mmwave.

\bibitem{b05} V. Raghavan, M. Chi, M. A. Tassoudji, O. H. Koymen and J. Li, “Antenna Placement and Performance Tradeoffs With Hand Blockage in Millimeter Wave Systems,” in \textit{IEEE Transactions on Communications}, vol. 67, no. 4, pp. 3082-3096, Apr. 2019.

\bibitem{b06} F. Abinader, C. Rom, K. Pedersen, S. Hailu and N. Kolehmainen, “System-Level Analysis of mmWave 5G Systems with Different Multi-Panel Antenna Device Models,” \textit{2021 IEEE 93rd Vehicular Technology Conference (VTC2021-Spring)}, 2021, pp. 1-6.

\bibitem{b07} I. Viering, B. Wegmann, A. Lobinger, A. Awada and H. Martikainen, "Mobility Robustness Optimization Beyond Doppler Effect and WSS Assumption," \textit{2011 8th International Symposium on Wireless Communication Systems}, 2011, pp. 186-191,

\bibitem{b08} 3GPP, \textit{Feature lead summary\#3 of Enhancements on Multi-beam Operations}, document R1-1907860, 3GPP TSG RAN WG1 Meeting \#97, LG Electronics, Reno, USA, May 2019.

\bibitem{b09} U. Karabulut, A. Awada, I. Viering, A. N. Barreto and G. P. Fettweis, “Low Complexity Channel Model for Mobility Investigations in 5G Networks,” \textit{2020 IEEE Wireless Communications and Networking Conference (WCNC)}, 2020, pp. 1-8.

\bibitem{b10} U. B. Elmali, A. Awada, U. Karabulut and I. Viering, “Analysis and Performance of Beam Management in 5G Networks,” \textit{2019 IEEE 30th Annual International Symposium on Personal, Indoor and Mobile Radio Communications (PIMRC)}, 2019, pp. 1-7.

\bibitem{b11}A. Ali et al., “System Model for Average Downlink SINR in 5G Multi-Beam Networks,” \textit{2019 IEEE 30th Annual International Symposium on Personal, Indoor and Mobile Radio Communications (PIMRC)}, 2019, pp. 1-6.

\bibitem{b12} 3GPP, “NR; NR and NG-RAN Overall Description; Stage 2,” Technical Report (TR) 38.300 V 16.4.0, 3rd Generation Partnership Project (3GPP), Dec. 2020. 

\bibitem{b13} 3GPP, “Study on New Radio Access Technology Physical Layer Aspects,” Technical Report (TR) 38.802 V14.2.0, 3rd Generation Partnership Project (3GPP), Sep. 2020.

\bibitem{b14} 3GPP, “Requirements for Support of Radio Resource Management,” Technical Report (TR) 38.133 V16.3.0, 3rd Generation Partnership Project (3GPP), Mar. 2020. 

\bibitem{b15} 3GPP, “Study on Channel Model for Frequencies from 0.5 to 100 GHz,” Technical Report (TR) 38.901 V15.0.0, 3rd Generation Partnership Project (3GPP), Jun. 2018.

\bibitem{b17} F. B. Tesema, A. Awada, I. Viering, M. Simsek and G. Fettweis, “Evaluation of Context-Aware Mobility Robustness Optimization and Multi-Connectivity in Intra-Frequency 5G Ultra Dense Networks,” in \textit{IEEE Wireless Communications Letters}, vol. 5, no. 6, pp. 608-611, Dec. 2016.

\end{thebibliography}
\end{document}